\documentclass[cits]{PoS}  

\usepackage[utf8]{inputenc}
\usepackage{amsmath}
\usepackage{amssymb}  
\usepackage{mathrsfs} 
\usepackage[retainorgcmds]{IEEEtrantools} 
\usepackage{multirow}
\usepackage{xspace}
\usepackage{lineno}

\usepackage{datetime} 

\usepackage{epstopdf}
\DeclareGraphicsExtensions{.jpg,.eps,.png,.pdf}

\usepackage{textpos}  
\usepackage{epigraph} 
\renewcommand\descriptionlabel[1]{\hspace\labelsep\normalfont\itshape #1 :}

{ \begin{list}%
        {$\bullet$}%
        {\setlength{\labelwidth}{30pt}%
         \setlength{\leftmargin}{35pt}%
         \setlength{\itemsep}{\parsep}}}%
{ \end{list} }

\definecolor{DarkGray}{RGB}{60,60,60}
\definecolor{LightGray}{RGB}{145,145,145}

\definecolor{Sepia}{RGB}{94,38,18}    
\definecolor{IndianRed}{RGB}{176,23,31}
\definecolor{OrangeRed4}{RGB}{139,37,0}
\definecolor{DarkRed}{RGB}{139,0,0}

\definecolor{DarkSlateBlue}{RGB}{72,61,139}
\definecolor{Cobalt}{RGB}{61,89,171}
\definecolor{RoyalBlue4}{RGB}{39,64,139}
\definecolor{DodgerBlue4}{RGB}{16,78,139}
\definecolor{SteelBlue4}{RGB}{54,100,139}
\definecolor{DeepSkyBlue4}{RGB}{0,104,139}

\definecolor{LightGreen}{RGB}{0,200,0}

\usepackage[colorinlistoftodos]{todonotes} 

\usepackage[utf8]{inputenc}

\newcommand{\ie}        {$i.e.$}
\newcommand{\eg}        {$e.g.$}
\newcommand{\fig}       {\textsc{f}ig.~}
\newcommand{\tab}       {\textsc{t}ab.~}

\newcommand {\pT}        {\ensuremath{p_{\mathrm{\textsc{t}}}}}
\newcommand {\meanpT}    {\ensuremath{\langle p_{\mathrm{T}} \kern-0.1em\rangle}}
\newcommand {\mean}[1]   {\ensuremath{\langle #1 \kern-0.1em\rangle}} 
\newcommand {\sqrtSnn}   {\ensuremath{\sqrt{s_{\textsc{nn}}}}}
\newcommand {\sqrtS}     {\ensuremath{\sqrt{s}}}

\newcommand {\rapJpsi}   {\mbox{$y_{\tiny \rmJpsi}$}}

\newcommand {\cTau}      {\ensuremath{c.\tau}}

\newcommand {\dNsquared} {\ensuremath{\mathrm{d^{2}}N }}
\newcommand {\dsquared} {\ensuremath{\mathrm{d^{2}} }}
\newcommand {\dpt}       {\ensuremath{\mathrm{d}\pT }}
\newcommand {\dy}        {\ensuremath{\mathrm{d}y}}

\newcommand {\dNchdeta}  {\ensuremath{\mathrm{d}N_\mathrm{ch}/\mathrm{d}\eta }}

\newcommand {\Raa}       {\ensuremath{R_\mathrm{AA}}}

\newcommand {\pp}        {\mbox{$\mathrm {p\kern-0.05em p}$}}
\newcommand {\ppBoldMath} {\mbox{$\mathrm { \mathbf p\kern-0.05em \mathbf p }$}}
\newcommand {\ppbar}     {\mbox{$\mathrm {p\overline{p}}$}}
\newcommand {\PbPb}      {\ensuremath{\mbox{Pb--Pb} }}
\newcommand {\AuAu}      {\ensuremath{\mbox{Au--Au} }}

\renewcommand {\AA}      {\ensuremath{\mbox{A--A}}}
\newcommand {\pA}        {\ensuremath{\mbox{p--A}}}

\newcommand {\hPM}       {\ensuremath{h^{\pm}}}

\newcommand {\MeanNpart} {\mbox{\ensuremath{< \kern-0.15em N_{part} \kern-0.15em >}}}

\newcommand {\Lint}      {\ensuremath{L_\mathrm{int}}}

\newcommand {\mass}     {\mbox{\rm MeV$\kern-0.15em /\kern-0.12em c^2$}}
\newcommand {\tev}      {\mbox{${\rm TeV}$}}

\newcommand {\mmom}     {\mbox{\rm MeV$\kern-0.15em /\kern-0.12em c$}}
\newcommand {\gmom}     {\mbox{\rm GeV$\kern-0.15em /\kern-0.12em c$}}
\newcommand {\kmass}    {\mbox{\rm keV$\kern-0.15em /\kern-0.12em c^2$}}
\newcommand {\mmass}    {\mbox{\rm MeV$\kern-0.15em /\kern-0.12em c^2$}}
\newcommand {\gmass}    {\mbox{\rm GeV$\kern-0.15em /\kern-0.12em c^2$}}

\newcommand {\mim}      {\mbox{$ \mu {\rm m}$}}

\newcommand {\invmub}   {\mbox{$\mu{\rm b}^{-1}$}}

\newcommand {\dg}       {\mbox{$\kern+0.1em ^\circ$}}

\newcommand{\rmLambdas}         {\mbox{$\mathrm {\Lambda \kern-0.2em + \kern-0.2em \overline{\Lambda}}$}}

\newcommand{\rmXiPM} { \mbox{$\kern-0.2em \mathrm{\vcenter{\offinterlineskip \vskip-1.0ex\hbox{\tiny \kern-0.05em- \kern-0.3em- \kern-0.3em-} \vskip+0.1ex\hbox{$\Xi$}}^{\protect \tiny ^\fivedots \kern-0.75em _{\relbar} \kern-.7em}}$} }

\newcommand{\rmXis}     {\mbox{$\mathrm {\Xi^{-} \kern-0.3em + \kern-0.1em \overline{\Xi}^{+}}$}}

\newcommand{\rmOmegas}  {\mbox{$\mathrm {\Omega^{-} \kern-0.3em +  \kern-0.1em \overline{\Omega}^{+}}$}}
\newcommand{\rmOmegaPM} { \mbox{$\kern-0.2em \mathrm{\vcenter{\offinterlineskip \vskip-1.0ex\hbox{\tiny \kern-0.05em- \kern-0.3em- \kern-0.3em-} \vskip+0.1ex\hbox{$\Omega$}}^{\protect \tiny ^\fivedots \kern-0.7em _{\relbar} \kern-.7em}}$} }

\newcommand{\rmJpsi}    {\mbox{$\mathrm {J\kern-0.05em /\kern-0.05em\psi}$}}
\newcommand{\rmPsiTwoS} {\mbox{$\mathrm {\psi(2S)}$}}
\newcommand{\rmChicZero}{\mbox{$\mathrm {\chi_{c_0}}$}}
\newcommand{\rmChicOne} {\mbox{$\mathrm {\chi_{c_1}}$}}
\newcommand{\rmChicTwo} {\mbox{$\mathrm {\chi_{c_2}}$}}
\newcommand{\rmChicJ}   {\mbox{$\mathrm {\chi_{c_J}}$}}

\newcommand{\rmLambdaC} {\mbox{$\mathrm {\Lambda}_{c}^{+}$}}
\newcommand{\rmDzero}   {\mbox{$\mathrm {D}^{0}$}}
\newcommand{\rmDplus}   {\mbox{$\mathrm {D}^{+}$}}
\newcommand{\rmDstar}   {\mbox{$\mathrm {D}^{*}$}}
\newcommand{\rmDs}      {\mbox{$\mathrm {D}^{+}_{s}$}}

\newcommand{\rmBzero}   {\mbox{$\mathrm {B^{0}}$}}
\newcommand{\rmBplus}   {\mbox{$\mathrm {B^{+}}$}}

\newcommand{\rmBplusMinus}   {\mbox{$\mathrm {B^{\pm}}$}}
\newcommand{\rmBzeroS}  {\mbox{$\mathrm {B^{0}_s}$}}

\newcommand{\rmUpsOneS} {\mbox{$\mathrm {\Upsilon(1S)}$}}
\newcommand{\rmUpsTwoS} {\mbox{$\mathrm {\Upsilon(2S)}$}}
\newcommand{\rmUpsThreeS}  {\mbox{$\mathrm {\Upsilon(3S)}$}}

\newcommand{\ccbar}     {\mbox{$c\overline{c}$}}

\newcommand{\EplusEminus}  {\mbox{$\mathrm {e^+e^-}$}}
\newcommand{\MuPlusMuMinus}  {\mbox{$\mathrm {\mu^+\mu^-}$}}

\title{Measurements of inclusive \rmJpsi{} production in \PbPb{} collisions at \sqrtSnn{} = 2.76 TeV with the ALICE experiment}

\ShortTitle{Inclusive \rmJpsi{} production in \PbPb{} collisions at \sqrtSnn{} = 2.76 TeV with ALICE}

\author{\speaker{Antonin Maire}, for the ALICE Collaboration
         \\
        Physikalisches Institut - Heidelberg, Germany\\
        E-mail: \email{antonin.maire@cern.ch}}

\abstract{
Charmonium is a prominent probe of the Quark-Gluon Plasma (QGP), 
expected to be formed in ultrarelativistic heavy-ion (\AA) collisions. 
It has been predicted that the \rmJpsi{}(\ccbar) particle is dissolved in the deconfined medium created in \AA{} systems. 
However this suppression can be counterbalanced via regeneration of the charm/anti-charm bound state in QGP
or via statistical production at the phase boundary.
At LHC energies, the latter mechanisms are expected to play a more important role, due to a charm production cross section significantly larger than at lower energies.

Measurements obtained by the ALICE experiment for inclusive \rmJpsi{} production are shown, making use of \PbPb{} data at \mbox{\sqrtSnn{} = 2.76 TeV}, collected in 2010 and 2011. In particular, the focus is given on the nuclear modification factor, \Raa, derived for forward ($2.5 < y < 4$) and mid rapidities ($|y| < 0.9$), both down to zero transverse momentum (\pT). The centrality, $y$ and \pT{} dependences of \Raa{} are presented and discussed in the context of theoretical models, together with PHENIX and CMS results.

}

\FullConference{

\vspace{2.0cm}

X$^{th}$ Quark Confinement and the Hadron Spectrum\\
                 8–12 October 2012\\
                 TUM Campus Garching, Munich, Germany}

\begin{document}


\maketitle

%


\section{Introduction : studying \rmJpsi{} production in ultrarelativistic heavy-ion collisions}

Ultrarelativistic heavy-ion (\AA) collisions are expected to give access to a medium where quarks and gluons are deconfined, the so-called Quark-Gluon Plasma (QGP).
In order to test experimentally the deconfinement of the created medium, several signatures have been proposed. 
Among them, the melting of \rmJpsi{} within the QGP \cite{matsui1986jpsi}.
The idea is that colour screening in the medium prevents the bound states \ccbar{} to form. 
This results in a \rmJpsi{} production suppressed in \AA{} relative to the corresponding production in reference systems, like proton-proton (\pp) or proton-nucleus (\pA) collisions (hadronic collisions without QGP).
The idea developed for \rmJpsi{} can actually be extended to other charmonium and bottomonium states. The resulting suppression hierarchy is expected to depend on the binding energy of the considered quarkonium state. This leads to the notion of the \emph{sequential melting} \cite{mocsy2007quarkoniumMeltingTemp} : the states that are more weakly bound may dissociate in the deconfined medium from lower temperatures on ("$T_{disso}[\rmPsiTwoS]  \lesssim T_{disso}[\rmChicJ] < T_{disso}[\rmJpsi] \approx T_{disso}[\rmUpsTwoS] < T_{disso}[\rmUpsOneS]$"). 

However, with experiments moving towards higher \AA{} collision energies (\sqrtSnn), possible scenarios for recombination of $c$ with $\bar{c}$ into charmonia must be considered \cite{braunmunzinger2000thermalAspectsOfCharmoniumProd}.
Indeed, the total production of charm and anti-charm quarks becomes more and more abundant with \sqrtSnn{} and, if recombination may appear non-negligible already at RHIC energies, it is expected to be sizeable at the LHC.
There are basically two complementary approaches to describe the (re)generation of charmonium :
\begin{itemize}
    \item \emph{thermal models :}  charm quarks are thermalised in the QGP and quarkonia are produced at the phase boundary via statistical hadronisation. The implicit hypothesis is that \ccbar{} is fully dissolved in the medium; the possibility to bind $q$ with $\bar{q}$ can only happen at the last stage of the collision \cite{andronic2007JpsiGenerationAtPhaseBoundaryInAAColl}.
    \item \emph{transport models :} the binding and dissociation of \ccbar{} take place directly in the QGP, the competition being described by a transport equation. In other words, a partial survival of charmonium is complemented by a continuous regeneration \cite{thews2001enhancedJpsiProdInDeconfMatter, chen2012meanFieldEffectOnJpsiProdInAAatRHICandLHC, zhao2011mediumModifAndProdCharmoniaAtLHC}.
\end{itemize}


\section{Data analysis and experimental remarks}

\subsection{Detector setup and data collection}

The \rmJpsi{} analyses presented in these proceedings make use of both di-lepton decay channels of \rmJpsi{} (see \tab \ref{tab:Maire-inclJpsiDecayChannel}).
The analysis at mid rapidity ($|\rapJpsi| < 0.9$) is based on the \EplusEminus{} channel. It hinges on the central-barrel detectors of the ALICE experiment \cite{alice2008pstationExpmtInJINST} : 
the Inner Tracking System (ITS) and Time Projection Chamber (TPC), for tracking 
and the TPC and the Time Of Flight detector (TOF), to identify electrons and partially remove contamination by other particle species.
The studies at forward rapidity ($2.5 < \rapJpsi < 4$) profit from the \MuPlusMuMinus{} channel. 
They rely on detectors placed beyond an absorber, \ie{} muon-tracking and muon-trigger chambers.

In both cases, the analysed data samples are \PbPb{} collisions at \sqrtSnn{} =  2.76 \tev. 
For mid rapidity, this is a combination of data collected in 2010 ($\Lint \approx 2$ \invmub) 
and 2011 ($\Lint \approx 13$ \invmub, \ie{} about 60 \% of \PbPb{} data collected that year).
For forward rapidity, this is exclusively data taken in 2011 ($\Lint \approx 70$ \invmub).

\subsection{Remarks : low-\pT{} reach and \emph{inclusive} \rmJpsi{} measurement...}

It must be noted that the ALICE acceptance enables the \rmJpsi{} identification down to zero transverse momentum (\pT) in both rapidity ranges, \ie{} in the phase space region where the bulk of charmonium is produced. This is a unique feature of the experiment for what concerns \AA{} at the LHC.

 \begin{table}[!hbt]
         \begin{center}
                 \begin{tabular}{l|r @{~} l|cr @{~} l|lr @{.} l}
                \noalign{\smallskip}\hline \noalign{\smallskip}
                    ~ & \multicolumn{2}{c|}{Particles}  &   mass ($\gmass$)   &  \multicolumn{2}{c|}{\cTau{} or width}  & decay channel & \multicolumn{2}{c}{B.R.} \\
                \noalign{\smallskip} \hline\hline \noalign{\smallskip}
                   \multirow{6}*{\rotatebox{90}{\small $\leftarrow \emph{prompt} \rightarrow$}} & \multirow{2}*{\rmJpsi}  &  \multirow{2}*{(\ccbar)}  & \multirow{2}*{3.097} & \multirow{2}*{92.9} & \multirow{2}*{\kmass} & $\rmJpsi \rightarrow \EplusEminus$      & 5&94~\% \\
                                                         &                         &                               &                      &                     &                       & $\rmJpsi \rightarrow \MuPlusMuMinus$  & 5&93~\% \\
                \noalign{\smallskip} \cline{2-9} \noalign{\smallskip}
                    & \rmChicZero & (\ccbar) & 3.415 & 10.4 & \mmass & $\rmChicZero \rightarrow \rmJpsi(1S) \gamma$ &  1&17~\% \\
                    & \rmChicOne & (\ccbar)  & 3.511 & 0.86 & \mmass & $\rmChicOne  \rightarrow \rmJpsi(1S) \gamma$ & 34&4~~~\% \\
                    & \rmChicTwo & (\ccbar)  & 3.556 & 1.98 & \mmass & $\rmChicTwo  \rightarrow \rmJpsi(1S) \gamma$ & 19&5~~~\% \\
                    & \rmPsiTwoS & (\ccbar)  & 3.686 & 304  & \kmass & $\rmPsiTwoS \rightarrow \rmJpsi + anything $ & 59&5~~~\% \\
                \noalign{\smallskip}\hline \hline \noalign{\smallskip}
                    \multirow{3}*{\rotatebox{90}{\small \emph{non-prompt}}} & \rmBzero & ($d\bar{b}$)    & 5.280 & 455 & \mim & & \multicolumn{2}{c}{} \\
                                                          & \rmBplus & ($u\bar{b}$)    & 5.279 & 492 & \mim & $\rmBzero, \rmBplusMinus, \rmBzeroS \rightarrow \rmJpsi + anything$ & 1&16~\% \\
                                                          & \rmBzeroS & ($s\bar{b}$)   & 5.367 & 449 & \mim & & \multicolumn{2}{c}{} \\
                \noalign{\smallskip}\hline \noalign{\smallskip}

                 \end{tabular}
        \caption{Main characteristics of particles contributing to the \emph{inclusive} \rmJpsi{} signal \cite{ParticleDataGroup2012}. The table includes mass, decay length or resonance width, as well as considered decay channel and corresponding Branching Ratio (B.R.).}
        \label{tab:Maire-inclJpsiDecayChannel}
         \end{center}
 \end{table}

The measurements presented in the following are for \emph{inclusive} \rmJpsi, \ie{} not only for direct \rmJpsi{} production but also \rmJpsi{} fed down from decays of particles with higher mass. As presented in \tab \ref{tab:Maire-inclJpsiDecayChannel}, two components can be distinguished :
\begin{itemize}
    \item \emph{the prompt component} for which the production vertex cannot be separated experimentally from the primary interaction region; it includes \rmJpsi{} issued by prompt decays of \rmChicJ{} and \rmPsiTwoS, together with primary \rmJpsi;
    \item \emph{the non-prompt component}, originating from beauty mesons, with a decay length which is large enough to allow a possible discrimination between the decay vertex and the primary one. 
\end{itemize}

For an inclusive measurement, with a \rmJpsi{} momentum ranging from \pT{} = 0 to about 8 \gmom,
experimental data\footnotemark{} at \tev{} scale suggest that the inclusive \rmJpsi{} yield has the following relative contributions:
\mbox{35-55} \% of direct production, 25-35 \% from \rmChicJ, 10-15 \% from \rmPsiTwoS{} and 10-15 \% from B mesons.

\footnotetext{In pp at the LHC or \ppbar{} at the Tevatron; there is no equivalent picture measured in \AA.}


\section{Experimental results : nuclear modification factor}

The comparison of the \rmJpsi{} production in \AA{} and \pp{} is studied via the nuclear modification factor, \Raa, defined as follows :


\begin{equation}
    \Raa(\pT, \, y, \, \text{centrality class } \textcolor{gray}{i}) 
    = 
    \frac{
    \dNsquared^{AA}_{\textcolor{gray}{i}}({\small \rmJpsi}) / \dpt\dy}
    {\langle T^{AA} \rangle_{\textcolor{gray}{i}} 
    \quad 
    \dsquared\sigma^{\scriptsize \pp}({\small \rmJpsi}) / \dpt\dy}
\end{equation}

\noindent with the \rmJpsi{} yields in \PbPb{} at \sqrtSnn{} =  2.76 \tev{} as the basis of the present measurements.
The study can be performed differentially as functions of collision centrality, rapidity and/or momentum, depending on the collected data statistics.
The values of the nuclear overlap function $\langle T^{AA} \rangle_{\textcolor{gray}{i}}$ are derived from a Glauber Model, for each centrality interval, 
and correspond to a normalisation of \PbPb{} yields to the adapted number of \emph{binary} nucleon-nucleon collisions. 
As for the \pp{} reference, the inclusive \rmJpsi{} cross sections, $\sigma^{\scriptsize \pp}({\small \rmJpsi})$, are taken from \cite{alice2012JpsiInPP2pt76TeV}, measured also at \sqrtS{} = 2.76 \tev, for both mid and forward rapidity.

    \begin{figure}[!hbt]
        \centering
            \includegraphics[width=0.65\textwidth, angle=0, clip=true, trim=0cm 0 0 0]{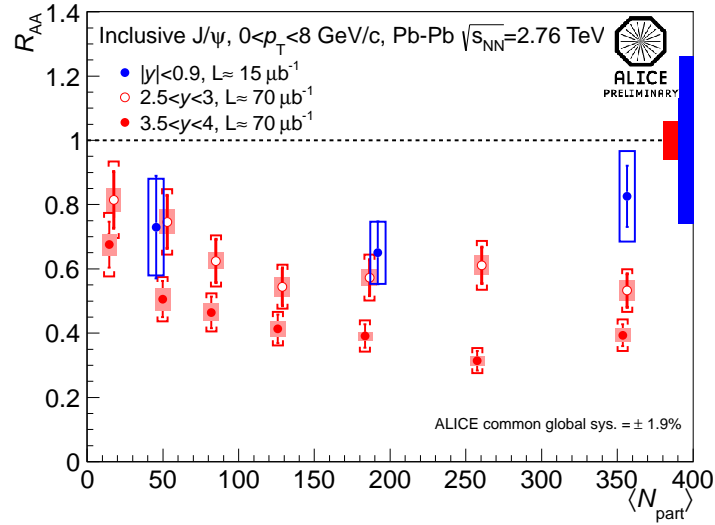}
        \caption{Nuclear modification factor \Raa{} for inclusive \rmJpsi{} as a function of collision centrality, 
            measured by the ALICE experiment in \PbPb{} collisions at \sqrtSnn{} = 2.76 \tev, in three rapidity intervals ranging from forward to mid rapidity. 
            Vertical error bars stand for statistical uncertainties; blue empty boxes, for systematic uncertainties; red brackets and shaded boxes, for partially correlated and uncorrelated systematic uncertainties, respectively. The filled boxes sitting at unity indicate the respective pp reference uncertainties.}
        \label{Fig:Maire-Fig1-ALICE-RaaInclJpsi-FwdAndMid}
    \end{figure}

The figure \ref{Fig:Maire-Fig1-ALICE-RaaInclJpsi-FwdAndMid} shows \Raa{} of inclusive \rmJpsi{} extracted for three rapidity ranges -- from most forward to mid $y$ -- as a function of centrality, embodied here by the average number of nucleons participating in a \PbPb{} collision, $\langle N_{part} \rangle$. 
One can observe a weak centrality dependence of the suppression, for each rapidity interval, 
and a suppression which seems less and less pronounced while going towards mid rapidity.
Note that, for the results derived at mid rapidity, the uncertainties remain rather significant and dominated by systematic uncertainties related to the \pp{} reference as well as the signal extraction in \PbPb.

    \begin{figure}[!htb]
        \begin{minipage}[c]{.48\linewidth}
                \includegraphics[width=1.05\textwidth, angle=0, clip=true, trim=0cm 0 0 0]{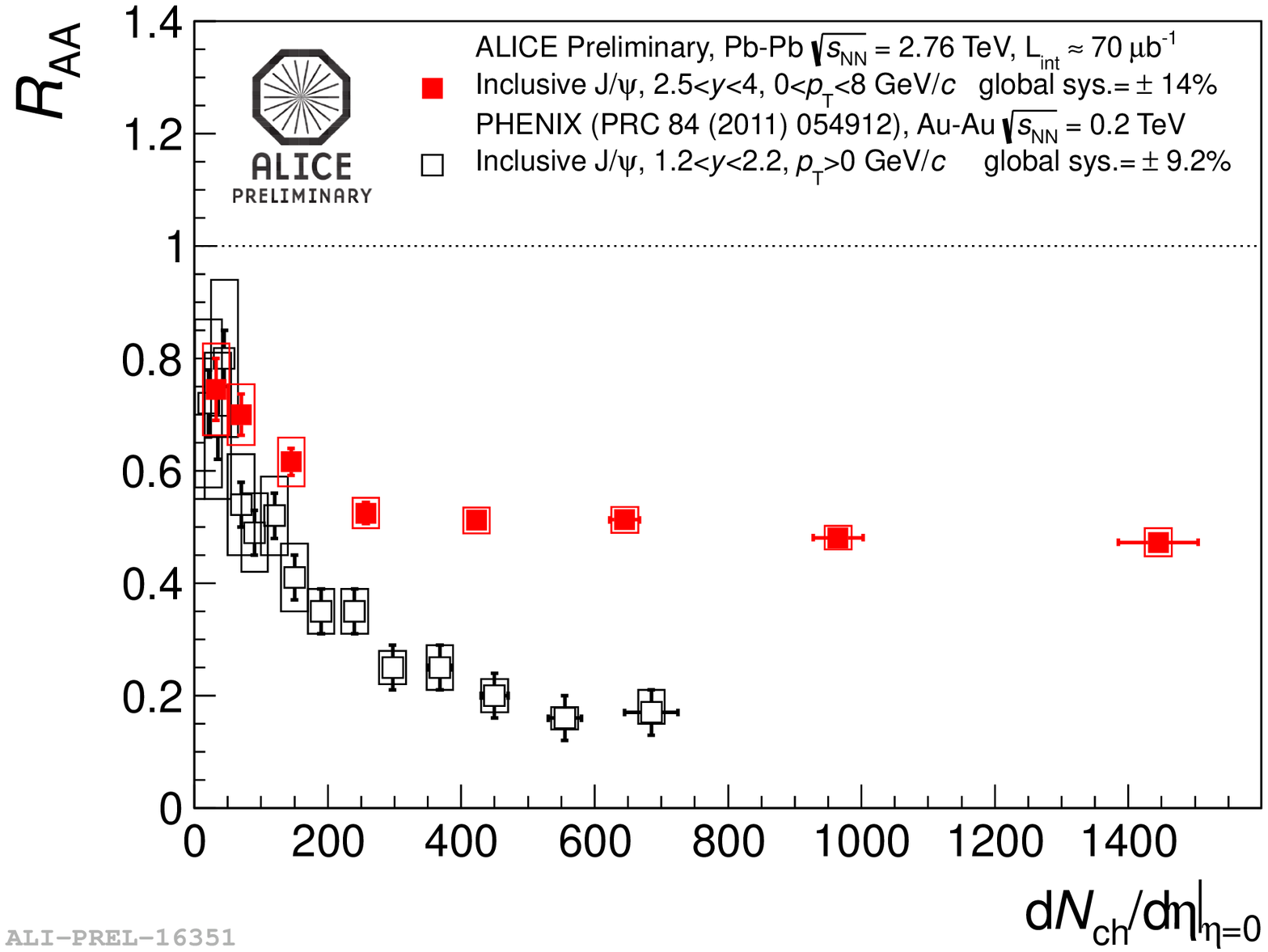}
                \hspace*{2.0cm} (\emph{a})
        \end{minipage} \hfill
        \begin{minipage}[c]{.48\linewidth}
                \includegraphics[width=1.05\textwidth, angle=0, clip=true, trim=0cm 0 0 0]{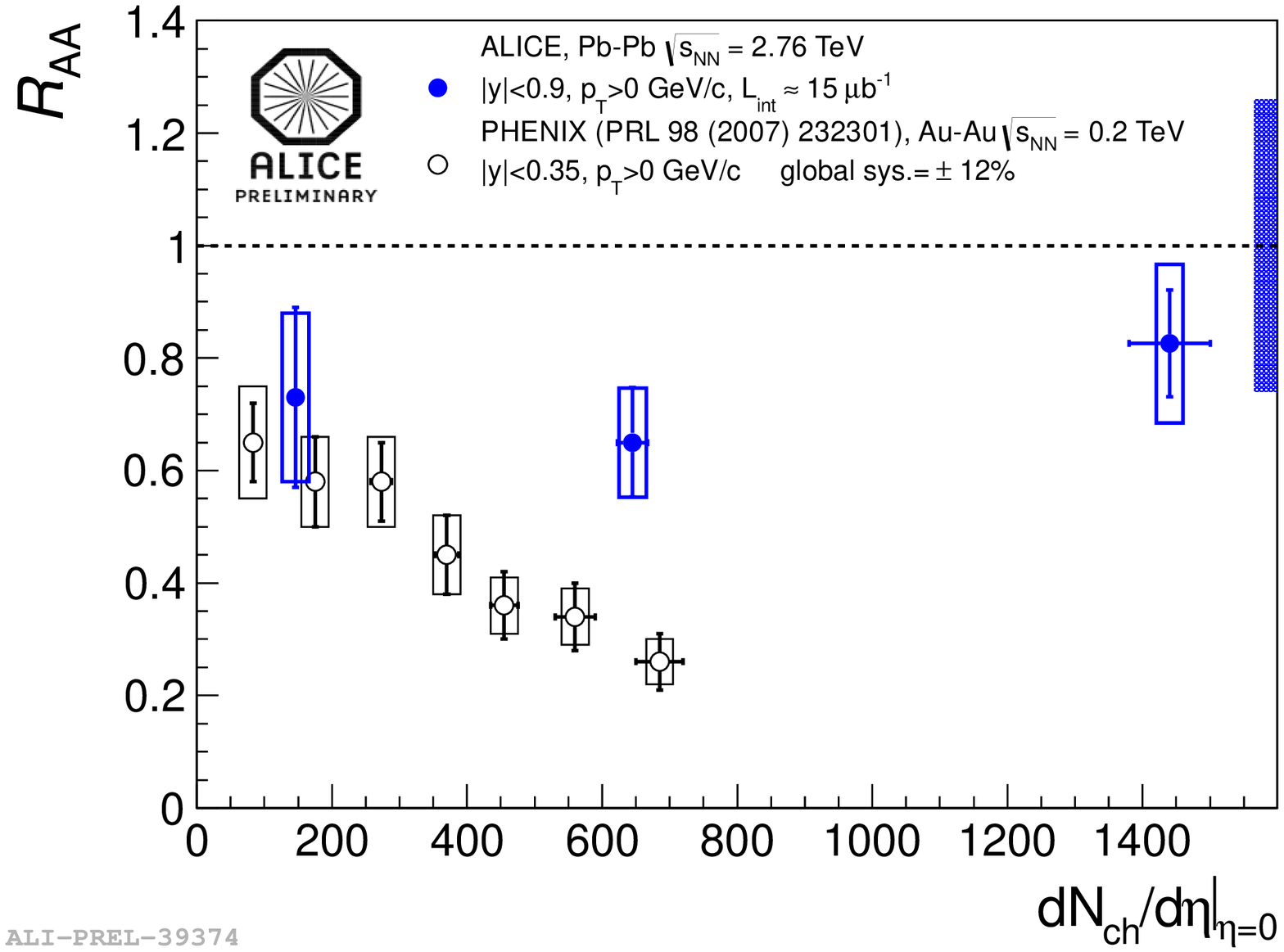}
                \hspace*{2.0cm} (\emph{b})
            \end{minipage} \hfill
                \caption{Centrality dependence of ALICE \Raa(inclusive \rmJpsi) at (\emph{a}) forward rapidity, compared with equivalent PHENIX results \cite{phenix2011JpsiRaaAtFwdRapInAuAu200GeV} and (\emph{b}) mid rapidity, compared with corresponding PHENIX results \cite{phenix2007JpsiRaaAtMidRapInAuAu200GeV}.}
                \label{Fig:Maire-Fig2-CompCentrality-ALICEvsPHENIX}
    \end{figure}

    \begin{figure}[!htb]
        \begin{minipage}[c]{.48\linewidth}
                \includegraphics[width=1.05\textwidth, angle=0, clip=true, trim=0cm 0 0 0]{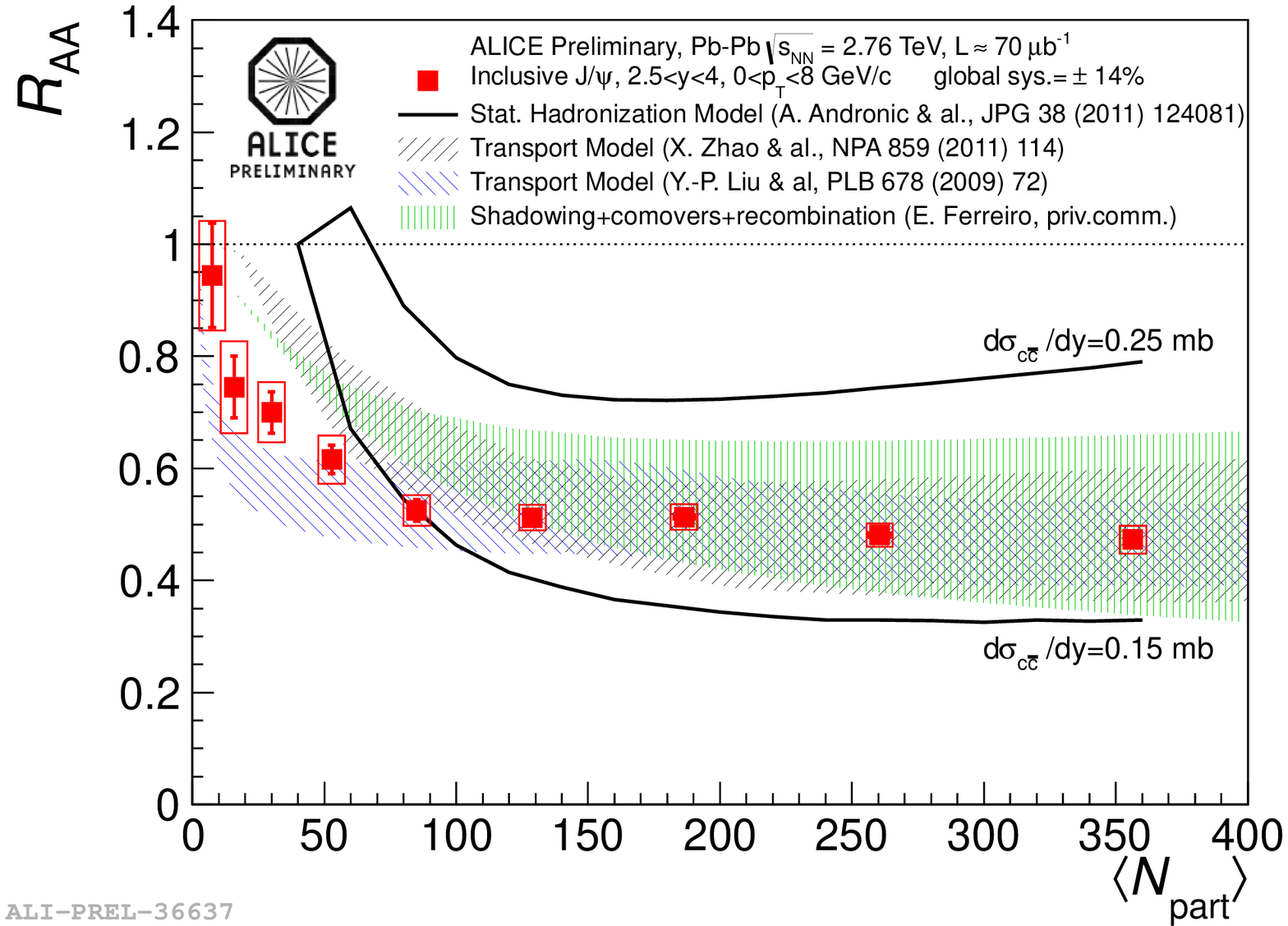} \\
                \hspace*{2.0cm} (\emph{a})
        \end{minipage} \hfill
        \begin{minipage}[c]{.48\linewidth}
                \includegraphics[width=1.05\textwidth, angle=0, clip=true, trim=0cm 0 0 0]{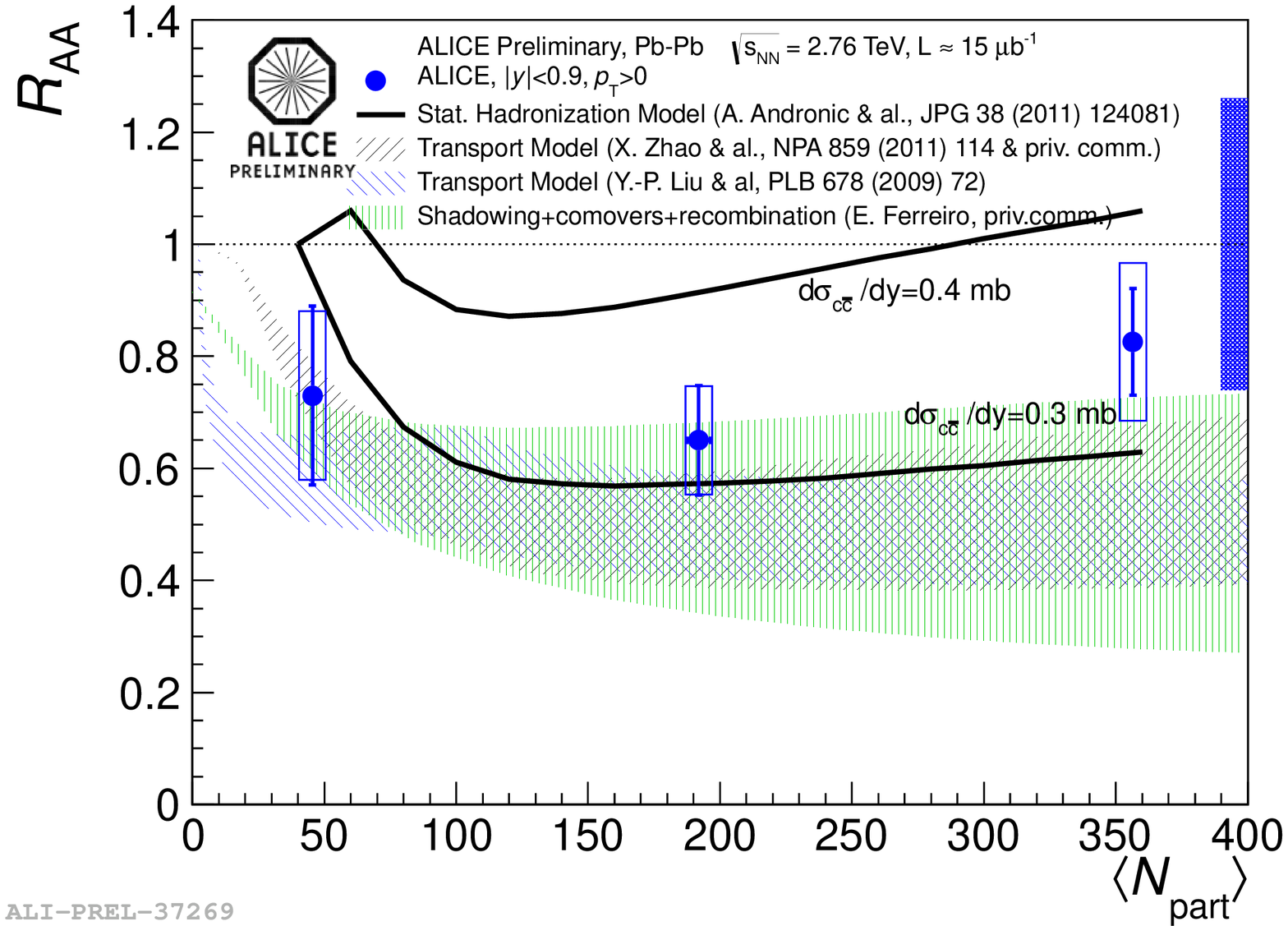} 
                \hspace*{2.0cm} (\emph{b})
            \end{minipage} \hfill
                \caption{ALICE centrality dependence for \Raa(inclusive \rmJpsi) compared with several models (thermal \cite{andronic2011thermalisationCharmQM2011}, transport \cite{zhao2011mediumModifAndProdCharmoniaAtLHC, liu2009JpsiRaaAtRHICandLHCByTransportModel-PLB} and co-movers \cite{ferreiro2012comoversJpsiAtLHC}), at (\emph{a}) forward rapidity and (\emph{b}) mid rapidity.}
                \label{Fig:Maire-Fig4-CompCentrality-ALICEvsModels}
    \end{figure}

    \begin{figure}[!t]
        \begin{minipage}[c]{.48\linewidth}
                \includegraphics[width=1.05\textwidth, angle=0, clip=true, trim=0cm 0 0 0]{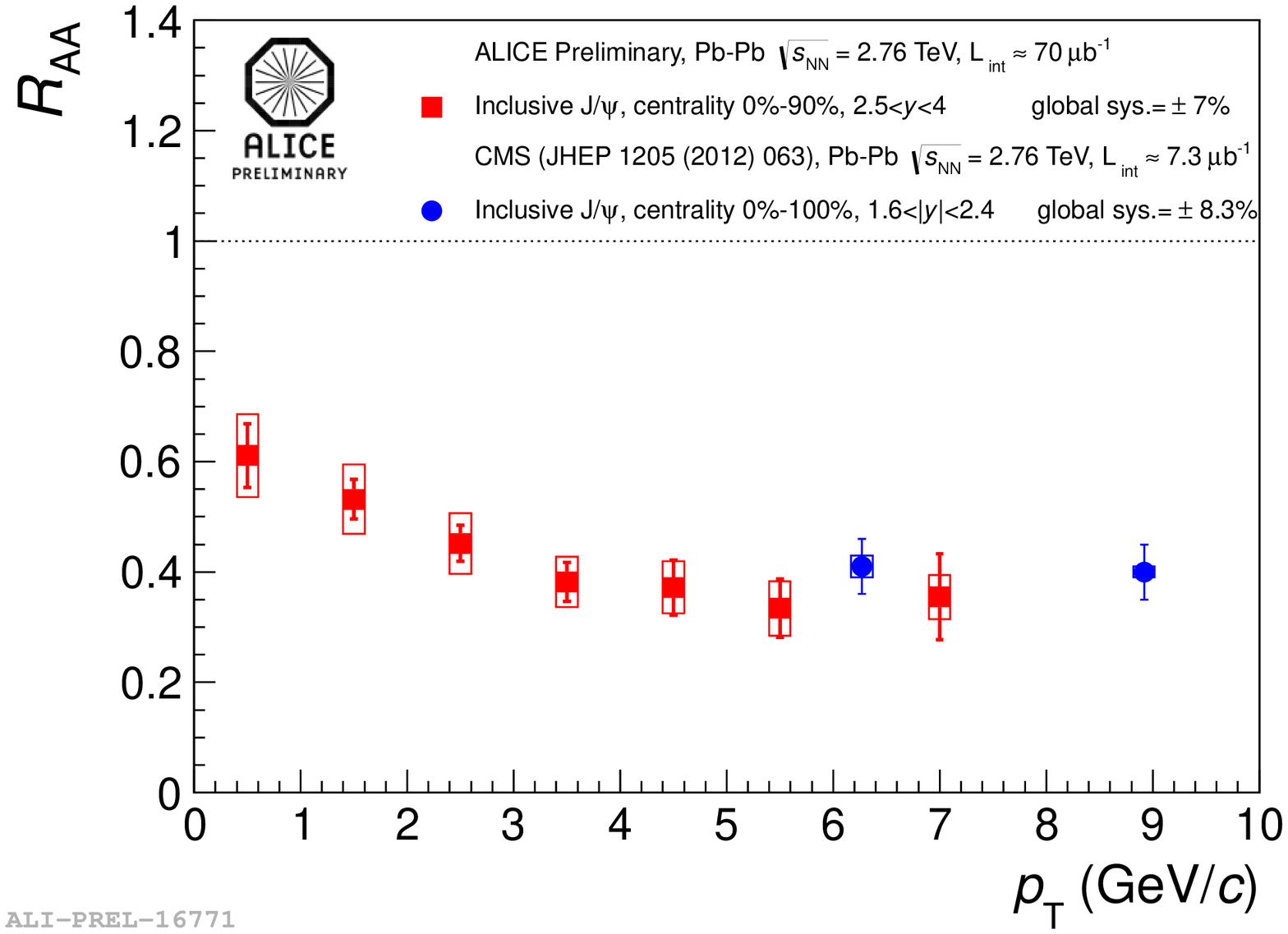}
                \hspace*{2.0cm} (\emph{a})
        \end{minipage} \hfill
        \begin{minipage}[c]{.48\linewidth}
                \includegraphics[width=1.05\textwidth, angle=0, clip=true, trim=0cm 0 0 0]{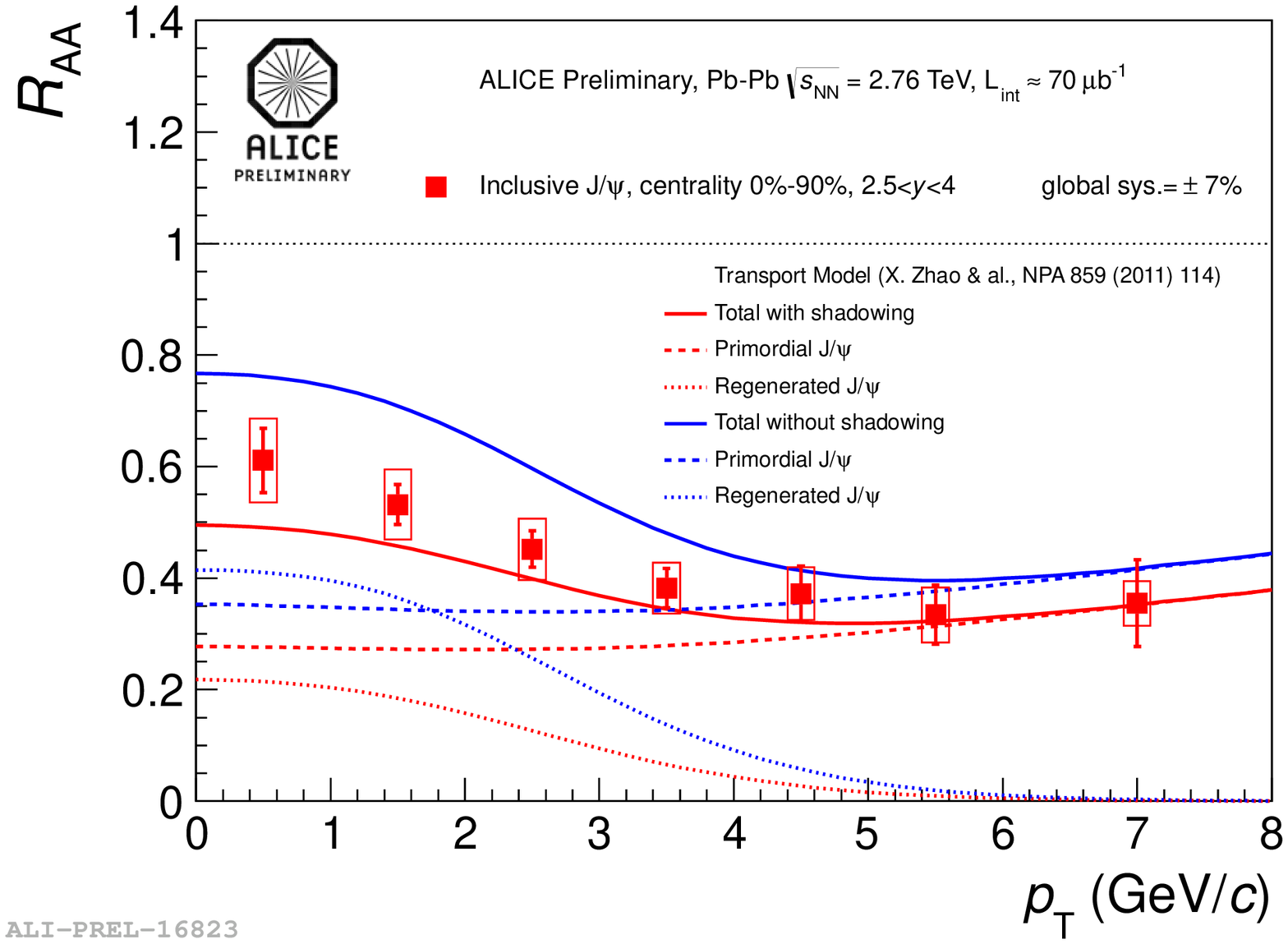}
                \hspace*{2.0cm} (\emph{b})
            \end{minipage} \hfill
                \caption{\pT{} dependence of ALICE \Raa(inclusive \rmJpsi) at forward rapidity, integrated over all collision centralities (\emph{a}) compared with CMS \cite{cms2012JpsiPromptAndNonPromptRAAInPbPb2pt76TeV} and (\emph{b}) compared to a transport model \cite{zhao2011mediumModifAndProdCharmoniaAtLHC}.}
                \label{Fig:Maire-Fig3-CompPtDepdce-ALICEfwdvsCMSandZhao}
    \end{figure}

The centrality dependence of \Raa{} in ALICE can be compared to PHENIX results obtained for \AuAu{} collisions at lower energies (\sqrtSnn{} =  0.2 \tev).
This is presented in \fig \ref{Fig:Maire-Fig2-CompCentrality-ALICEvsPHENIX}, 
where the charge particle density, \dNchdeta, at mid rapidity is used as centrality observable. 
The figure exhibits a centrality dependence which is clearly different between ALICE and PHENIX data; 
in particular, the 0.2 \tev{} \Raa{} in most central collisions appears noticeably lower, 
in each rapidity region, than the equivalent quantity at 2.76 \tev. 
This pattern is in contrast with the expectations based solely on the colour screening model, 
for which higher energy density (higher \dNchdeta) implies a stronger suppression.

This trend indicating less suppression at the LHC can be explained by several models, 
as plotted in \fig \ref{Fig:Maire-Fig4-CompCentrality-ALICEvsModels} : 
whether it deals with statistical hadronisation \cite{andronic2011thermalisationCharmQM2011}, 
transport  \cite{zhao2011mediumModifAndProdCharmoniaAtLHC, liu2009JpsiRaaAtRHICandLHCByTransportModel-PLB} 
or co-mover \cite{ferreiro2012comoversJpsiAtLHC} models, 
a fair agreement seems to be found between LHC data and phenomenological descriptions. 
It should be noticed that every model incorporates a significant fraction\footnotemark{} 
of $c$ and $\bar{c}$ recombination for most central collisions.
Any further interpretation and discrimination between models is precluded by uncertainties, on data, but also on models.
For instance, all of them are almost equivalently affected by the uncertainty of the total $c, \bar{c}$ production cross section, \ie{} the absence of accurate measurements covering open charm (\rmDzero, \rmDplus, \rmDstar, \rmDs{} mesons and \rmLambdaC{} baryon) and hidden charm (the whole charmonium family) over a wide region of phase space.

\footnotetext{By construction, it is 100~\% in the case of statistical models.}

With the current data, a last item to complement the description of \Raa(inclusive \rmJpsi) can be added though.
The figure \ref{Fig:Maire-Fig3-CompPtDepdce-ALICEfwdvsCMSandZhao} displays the \pT{} dependence of the nuclear modification factor,
for the forward rapidity region, in a wide centrality interval (0-90\%). 
The resulting differential measurements span from 0 to 8 \gmom{} 
and overlap at high \pT{} with the corresponding result (under similar centrality and rapidity selections) by the CMS collaboration \cite{cms2012JpsiPromptAndNonPromptRAAInPbPb2pt76TeV} (\fig \ref{Fig:Maire-Fig3-CompPtDepdce-ALICEfwdvsCMSandZhao}~\emph{a}).
Looking at the detailed comparison with one transport approach \cite{zhao2011mediumModifAndProdCharmoniaAtLHC} (\fig \ref{Fig:Maire-Fig3-CompPtDepdce-ALICEfwdvsCMSandZhao}~\emph{b}), data seem to be correctly reproduced by theory.
Based on the given model, the agreement suggests that \rmJpsi{} regeneration phenomena may essentially vanish from 5-6 \gmom{} on, 
whereas they should account for a sizeable contribution ($\approx 50$ \%) for \pT~$<$~3~\gmom.
Considering recombination as a mechanism preferentially working at low momentum, this emphasizes the importance of measuring \rmJpsi{} down to \pT{} = 0 to test this kind of prediction.


\section{Final considerations : complementing the picture}

While testing the original idea by Matsui and Satz of \rmJpsi{} suppression in QGP, 
the study of \rmJpsi{} in \AA{} collisions at LHC energies may be examined from two other angles.
 
On the one hand, at high \pT{}, \rmJpsi{} appears as the "hidden charm" piece in the puzzle of identified-hadron \Raa. 
Can we confirm or invalidate with proper accuracy a high-\pT{} picture like :\\
\vspace{-1.2cm}

\begin{equation}
\Raa[\hPM] < \Raa[D] \approx \Raa[\text{prompt \rmJpsi}] \approx \Raa[\rmPsiTwoS] < \Raa[B \rightarrow \rmJpsi] < \Raa[\rmUpsOneS] \nonumber
\end{equation}

\noindent The question is the flavour dependence and hierarchy of the suppression for particles at high momentum ($\pT > 6$ \gmom), ranging from 
gluons (\Raa[\hPM], \cite{alice2012unidentifiedRAA}) 
up to the hidden beauty (\Raa[\rmUpsOneS, \rmUpsTwoS, \rmUpsThreeS], \cite{cms2012seqUpsilonSupprPRL}) 
via open charm (\Raa[D], \cite{alice2012RaaForDmesonsInPbPb2pt76TeV}) 
and open beauty (\Raa[\rmJpsi{} from B], \cite{cms2012JpsiPromptAndNonPromptRAAInPbPb2pt76TeV}).

On the other hand, at low \pT, studies may address the question related to thermalisation of charm quarks.
On this aspect, the discussion must be enriched by \Raa{} measurements of other charmonium (\eg{} \rmPsiTwoS{} \cite{arnaldi2012JpsiAndPsi2SatFwdRapInPbPb2pt76TeV-QM2012}) together with measurement of elliptic flow of charm species.
In that sense, the ALICE Preliminary measurement indicating a non-zero elliptic flow 
(significance close to 3-$\sigma$ in 20-40 \% \PbPb{} collisions with \pT(\rmJpsi) $>$ 1.5 \gmom{} \cite{yang2012ellipticFlowForJpsiInPbPb2pt76TeV-QM2012}) 
is certainly a prominent result that will be closely followed in the future.\\

Last but not least, one should conclude the discussion with the relevance of \rmJpsi{} measurements in \pA{} collisions.
Nothing has been mentioned so far on the studies related to this system whereas \pA{} collisions define a crucial experimental milestone on the way going from \pp{} to \AA{} collisions.
There are \emph{non-QGP} effects foreseen in \AA{} collisions : parton distribution functions in nuclei can be modified (lowering of the \rmJpsi{} production probability due to parton shadowing in the nucleus), incident parton can loose energy in nucleus before the hard scattering, ...
All these phenomena can be labelled as \emph{Cold Nuclear Matter} (CNM) effects.

In RHIC data, CNM effects are expected to explain for a large part the inclusive \rmJpsi{} suppression
seen in most central \AuAu{} collisions at \sqrtSnn{} = 0.2 \tev{} 
(see for instance \fig 87 and 88, p. 126 of \cite{brambilia2011heavyQuarkoniumReview}).
At LHC energies, the same phenomena may account for a less dramatic effect but this has to be confirmed experimentally.
For that purpose, results from \pA{} data taking occurring at the LHC early 2013 will certainly act as
as the keystone of the \rmJpsi{} picture currently emerging with \AA{} data at \tev{} scale.



\bibliographystyle{myJHEP}


\phantomsection 

\addcontentsline{toc}{chapter}{References} 
\bibliography{aProceedingsQCHS_AntoninMaire_Biblio}

\end{document}